# Analyzing the Social Structure and Dynamics of E-mail and Spam in Massive Backbone Internet Traffic

Technical Report no. 2010-03

Farnaz Moradi, Tomas Olovsson, Philippas Tsigas
Computer Science and Engineering
Chalmers University of Technology
Göteborg, Sweden
{moradi, tomasol, tsigas}@chalmers.se

## ABSTRACT

E-mail is probably the most popular application on the Internet, with everyday business and personal communications dependent on it. Spam or unsolicited e-mail has been estimated to cost businesses significant amounts of money. However, our understanding of the network-level behavior of legitimate e-mail traffic and how it differs from spam traffic is limited. In this study, we have passively captured SMTP packets from a 10 Gbit/s Internet backbone link to construct a social network of e-mail users based on their exchanged e-mails. The focus of this paper is on the graph metrics indicating various structural properties of e-mail networks and how they evolve over time. This study also looks into the differences in the structural and temporal characteristics of spam and non-spam networks. Our analysis on the collected data allows us to show several differences between the behavior of spam and legitimate e-mail traffic, which can help us to understand the behavior of spammers and give us the knowledge to statistically model spam traffic on the network-level in order to complement current spam detection techniques.

## 1. INTRODUCTION

E-mail has an increasing role in human communications. In this paper, the social behavior and dynamics of e-mail traffic are studied. First we have created social e-mail networks based on SMTP traffic passively captured on a 10 Gbits/s backbone link of Swedish University network (SUNET).[1] An e-mail network is a graph of e-mail communications with user e-mail addresses as vertices and e-mail transmissions between them as edges. Then we have investigated the basic structural properties of these social networks by looking at the degree distribution, average path length, clustering coefficient, and strongly connected components for the constructed e-mail networks. Although it was previously believed that e-mail networks are scale-free [1, 2, 3, 4, 5, 6], we have shown that it is only valid for legitimate e-mail transmissions. By studying the graph of e-mail communications on the network-level, it can be seen that e-mail networks are not scale-free due to the large quantity of unsolicited traffic (also known as *spam*) on the Internet. To the best of our knowledge, this is the first large scale study of social properties of e-mail networks.

Study of the topology and structure of complex networks such as the Internet and World Wide Web, has attracted a lot of attention in order to answer basic questions such as "What does the Internet look like?" Analyses have shown that these social networks have fundamental structural differences from other types of networks such as random networks [7, 8]. Watts and Strogatz [9] showed that social networks are "small world" networks, which means that they are highly clustered and have short characteristic path lengths. The www graph [10, 11, 12], phone call graphs [13], online social networks [14], and movie actors network [15] are a few examples of such networks.

The unregulated growth of e-mail communications leads to a huge and complex social network. Ebel et al. [1] showed that e-mail networks, analogous to other social networks, show the characteristics of small world networks. They also showed that since the vertex connectivity of e-mail networks follows a power-law distribution, these networks exhibit the characteristics of scale-free networks as well [15]. The e-mail network studied in [1] was limited to a small number of student e-mail communications logged by the local mail server at their University. Although it is impossible to create complete global graph of e-mail communications, due to its large and increasing size, but in this paper we have studied significantly large e-mail networks that are not restricted to a single domain.

E-mail networks can be studied as both directed and undirected graphs. An undirected graph only represents the existence of a relationship between two e-mail addresses and is independent of who the sender or receiver of the e-mail is. However, in directed graphs the direction of the relationship is also taken into account.

Although current anti-spam tools are very efficient in hiding spam e-mails from users' mailboxes, the problem with unwanted traffic, waste of mail server resources, and false positives (legitimate e-mails incorrectly filtered as spam) caused by inaccurate filtering tools still remains untackled. In order to improve the defense against spam it is necessary to understand its behavior and detect deterministic characteristics that are indicative of spam traffic on network-level as close to source of spam as possible.

---
[1] http://vision.sunet.se

To better understand the dynamics and behavior of spam and to study the differences with legitimate e-mail, we have constructed two distinctive e-mail networks from spam traffic and legitimate e-mail transmissions (also known as *ham*). The e-mails have been classified as spam and ham by deploying a well-trained spam filtering tool (SpamAssassin [16]). This provided us a ground-truth and allowed us to study characteristics of spam traffic and how it differs from ham. Since many of the attempts to send spam get rejected by the pre-filtering strategies deployed on receiving mail servers (e.g., blacklisting, greylisting, DNS look-ups), a third e-mail network was generated from both delivered and rejected spam communications.

Our contributions in this paper are as follows:

- We have performed structural analysis on the e-mail networks generated from large amounts of network traffic in order to understand their structural characteristics.

- We have conducted a temporal analysis of e-mail networks and studied the effect of selecting different time windows on the structural properties that we consider.

- We have found surprising results in our study that although e-mail networks are considered scale-free in the literature, this is not the case.

- We have analyzed and compared the structural properties of legitimate e-mail (ham) networks and spam networks.

- We have observed properties of e-mail traffic that are indicative of spam. We believe these features can potentially be used for creating new anti-spam tools on the network-level.

The remainder of this paper is organized as follows. Section 2 describes the background and related work in the area. Section 3 describes the measurement settings used for data collection and generation of e-mail networks. Section 4 describes various structural and temporal properties of e-mail networks and how they differ for spam and ham traffic. A discussion of our results is presented in section 5. Finally, section 6 concludes this paper.

## 2. BACKGROUND AND RELATED WORK

The problem of understanding the structure and properties of network systems such as the Internet [17], World Wide Web [10, 11, 12], phone call graphs [13], e-mail networks [1], interaction of users in mobile systems [18], and online social networks [14], have strongly attracted researchers. Studies of structural properties such as the "small world phenomenon", or scale-free behavior of networks, etc., have revealed that social networks are fundamentally different from other types of networks [7].

Complementary, statistical analysis of e-mail traffic in order to model the e-mail workload, or study of deterministic properties of spam and ham, have also been of great interest during the last few years. Understanding the distinguishing characteristics of spam traffic is necessary for development of new mechanisms that can keep up with its dynamic behavior.

Recent attempts to characterize and analyze e-mail traffic have used different types of e-mail datasets such as SMTP log files of mail servers [19], spam e-mails captured in honeypots or relay sinkholes [20], flow-level data collected by network gateway routers [21], and e-mail traffic captured on the network-level [22].

The first study of the topology of e-mail networks was presented in Ebel et al. [1]. They studied an e-mail network generated from log files of the mail server of their university (Kiel University) over a period of 112 days. The network contained 59,812 nodes with average degree 2.28. They showed that this e-mail network exhibits a scale-free degree distribution and has small world properties. Our e-mail network is not limited to only one domain and is significantly larger (contains $5,971,825$ nodes and $10,610,875$ edges), although it was created over a shorter period of time.

Newman et al. [23] used an e-mail network that was generated from e-mail address books stored in a large University computer system, to study its structure and the potential of spreading computer viruses via e-mail. The network we have studied here is different from theirs, since our dataset comes from real e-mails that were actually transmitted during the measurement period and not from contact-lists/address books.

Caldarelli et al. [6] also studied e-mail networks in order to measure the strength of the relations between users, by keeping track of the number of e-mails that were received from a given sender. They analyzed data sets from five mailboxes containing e-mails from different numbers of senders during 3, 5 and 10 years. They showed that the degree distribution of all the e-mail networks are similar and follow a power-law distribution.

Use of social e-mail networks for distinguishing between spammers and legitimate e-mail senders was first proposed by Boykin et al. in [5]. They generated an e-mail graph from e-mail headers of one user's mailbox.

They showed that this network consists of several disconnected components and the spammers construct a large connected component with a low clustering coefficient. This property was used to generate blacklists of spammers and whitelists of legitimate users. Their approach is limited to individual users and does not scale for the whole network.

Gomes et al. [2] generated two different types of graphs from mail server log files of their University, and found graph theoretical metrics that structurally and dynamically differ for spam and ham e-mails. They investigated properties such as the in- and out-degree distribution, communication reciprocity, clustering coefficient, and the probability of visiting a node during a random walk. In our paper we have also studied some of these structural properties on our e-mail networks and have compared our observations when applicable.

In [24] Kong et al. suggested to use topological properties of e-mail networks to generate more efficient distributed collaborative spam filters. They made use of the very low percolation threshold caused by power-law degree distribution in the e-mail network studied in [1] to perform a percolation search and detect arrivals of previously seen spam on the network.

Brendel et al. [25] used e-mail networks to generate social relation pattern graphs for spammers and legitimate users. These pattern graphs were later used to detect typical and abnormal patterns on an e-mail network generated from mail server log files. Their algorithm was only efficient in recognizing normal users, but not adequate for detection of spammers.

In [26] Lam et al. extracted different structural features from e-mail social networks, such as in- and out-degree, communication reciprocity, and communication interaction average, and deployed them in building a learning-based spam detection method. Their e-mail network was generated from the publicly released Enron e-mail dataset. They simulated spam senders and injected spam e-mails into the dataset to study the effectiveness of their approach. A similar set of features was studied in [27] by Tseng et al. to construct a complete spam detection system based on an incremental support vector machine model. The dataset used to verify the efficiency of their system was also based on log files of a local university mail server (National Taiwan University).

Table 1 summarizes the properties of the e-mail networks described here. All of the above studies have taken place on relatively limited e-mail datasets. To the best of our knowledge, our study is the first study of social structure of e-mail networks on the Internet on a large scale.

## 3. MEASUREMENT SETTINGS

In this section, the methodology and system settings used to collect data and generate the e-mail networks are described.

We have constructed an e-mail network from SMTP packets captured on a backbone link of the Swedish University Network (SUNET) during a period of one week in March 2010. More than 400 million SMTP packets (filtered on port 25) were passively collected from a 10 Gbit/s link in both directions [28, 29], and were aggregated into more than 24 million flows to allow retrieval of complete e-mails. Around 12.5 million e-mails and 6.2 million mail server responses were extracted from these flows. The rest of the flows had no payload (scanning, incomplete, etc.).

Each flow carried one or more e-mails and we have classified each e-mail as *accepted* or *rejected* by the mail server or *incomplete*. Accepted e-mails (e-mails that are delivered by the receiving mail server) contain the basic SMTP commands ("MAIL FROM", "RCPT TO" and "DATA"), e-mail headers, e-mail body, and a line containing only a "." (period or full stop), which indicates the end of the e-mail data [30]. Rejected e-mails are those that cannot succeed to finish the SMTP command exchange phase and do not send any e-mail data. The rejection is generally due to the pre-filtering process deployed by receiving mail servers. Studies from inside a well-known spamming botnet [31, 32] have shown that the delivery success rate of spam is very low [31], and these unsuccessful attempts should therefore be considered as spam. Finally, incomplete SMTP flows are mainly due to scanning attempts and measurement equipment errors.

Furthermore, each accepted e-mail was classified as *spam* or *ham* to provide the ground-truth for further investigations on differences in their characteristics. Similar to [2, 33, 34], the classification was done by a well-trained SpamAssassin [16] filtering tool. After that, all the e-mail addresses were anonymized and e-mail contents were discarded in order to preserve privacy.

We have extracted e-mail addresses of the senders and the receivers from SMTP commands "MAIL FROM" and "RCPT TO", respectively. Each e-mail can have one sender and one or more receiver. E-mails with no senders (MAIL FROM:<>), which are mainly e-mails that are sent to notify about delivery failures, e-mails that have been rejected before sending these SMTP commands, and encrypted communications were ignored.

Each of the resulting e-mail networks were studied as both undirected and directed graphs $G(V, E)$, where each vertex $v \in V$ corresponds to an e-mail address and each edge $e \in E$ corresponds to an e-mail transmission (from the sender to the receiver in the directed graph).

We have studied the graphs generated for the data captured daily, and the graph generated from the whole week's traffic (weekly). Table 2 shows the statistical properties of the undirected and directed e-mail networks generated in each day and after the whole week. Mean node degree is defined as $\langle k \rangle = \sum_{v \in V} d(v)/|V|$, where $d(v)$ denotes the degree of node $v$. The directed e-mail network constructed after the seven days consists of $|V| = 6,096,959$ nodes and $|E| = 10,949,763$ directed edges with a mean degree $\langle k \rangle = 3.59$. This e-mail network is much larger than the networks studied previously [1, 2, 23, 6].

Table 1: Summary of datasets of related works

| Reference | $|V|$ | $|E|$ | Type | Dataset |
|---|---|---|---|---|
| Ebel et al. [1] (2002) | 59,812 | 86,130 | undirected/ directed | log files of the e-mail server at Kiel University |
| Gomes et al. [2] (2005) | 265,144 | 615,102 | directed | log files of the e-mail servers of a department in a university in Brazil |
| Newman et al. [23] (2002) | 16,881 | 57,029 | directed | address books on a university computer system server |
| Boykin et al. [5] (2005) | - | - | undirected | headers of e-mail messages in one user's inbox |
| Lam et al. [26] (2007) | 9,150 | - | directed | Enron dataset** and simulated spam accounts and e-mails |
| Brendel et al. [25] (2008) | - | 10,000 | directed | incoming e-mails to a faculty in Gdansk University |
| Kong et al. [24] (2006) | 56,969 | 84,190 | undirected | SCC of the dataset used in Ebel et al. [1] |
| Tseng et al. [27] (2009) | 637,064 | 2,865,633 | directed | e-mail server of the computer center in National Taiwan University |
| Caldarelli et al. [6] (2004) | 113-516* | 5,628-21,782* | undirected | five e-mail directories coming from authors' and colleagues' accounts |

* minimum and maximum number of nodes/edges in 5 different networks
** http://www.isi.edu/adibi/Enron/Enron Dataset Report.pdf

Table 2: Daily and weekly e-mail network statistics

| Day | $|V|$ | $|E|$ undirected | $|E|$ directed | $\langle k \rangle$ |
|---|---|---|---|---|
| 1 | 1,688,020 | 2,239,560 | 2,241,384 | 2.65 |
| 2 | 1,419,253 | 1,823,910 | 1,825,486 | 2.57 |
| 3 | 1,332,001 | 1,803,693 | 1,805,377 | 2.71 |
| 4 | 1,256,242 | 1606318 | 1,607,891 | 2.56 |
| 5 | 1,128,340 | 1,202,960 | 1,203,961 | 2.13 |
| 6 | 995,150 | 1,085,001 | 1,085,341 | 2.18 |
| 7 | 1,286,645 | 1,865,930 | 1,866,645 | 2.90 |
| Week | 6,096,959 | 10,949,763 | 10,962,143 | 3.59 |

In order to study the distinguishing properties of ham and spam e-mails, we have also generated two e-mail networks, one containing only legitimate e-mails as edges and another with spam e-mails as edges (according to the classification done by SpamAssassin). A third e-mail network was created from both delivered and rejected spam e-mail traffic. Due to existence of alternative links, only a small fraction of the captured SMTP flows are symmetric (both e-mail transmission and e-mail server responses are available), so it is not possible for us to verify the rejection reason for all e-mail communications. On the other hand, the error codes used in SMTP replies when rejecting a communication are not always informative of the rejection reason. However, previous studies (e.g., [21]) have shown that rejections are mainly because of spam pre-filtering strategies deployed by e-mail servers including blacklisting, greylisting, DNS lookups, and user database checks.

In general, the dataset used in this paper has the following characteristics:

- The e-mail traffic was captured on an Internet backbone link in both directions.

- The dataset does not contain internal e-mail communications (between internal users with each other) and is based on what can be seen on the backbone link (different from previously studied datasets [1, 2]).

- Multiple e-mail transmissions between two e-mail addresses are considered as one link for our anal-

ysis, so the generated graph does not contain duplicate edges (similar to [13]).

- An edge in the ham/spam/rejected network exists if a legitimate/spam/rejected e-mail was transmitted between the vertices connected by that edge (different than [2], which classified nodes as spammer or non-spammer to generate the networks).

- Self-loops exist in the graph representing e-mails sent by a user to herself.

Table 3: Daily and weekly ham network statistics

| Day | $|V|$ | $|E|$ | $\langle k \rangle$ |
|---|---|---|---|
| 1 | 146,914 | 120,016 | 1.63 |
| 2 | 164,747 | 145,538 | 1.77 |
| 3 | 132,496 | 110,836 | 1.67 |
| 4 | 125,327 | 104,062 | 1.66 |
| 5 | 106,470 | 86,731 | 1.63 |
| 6 | 57,862 | 46,672 | 1.61 |
| 7 | 73,037 | 54,136 | 1.48 |
| Week | 509,728 | 569,126 | 2.23 |

Table 4: Daily and weekly spam network statistics

| Day | $|V|$ | $|E|$ | $\langle k \rangle$ |
|---|---|---|---|
| 1 | 219,084 | 252,044 | 2.30 |
| 2 | 221,943 | 223,041 | 2.01 |
| 3 | 196,229 | 207,895 | 2.12 |
| 4 | 171,555 | 180,113 | 2.10 |
| 5 | 189,105 | 186,920 | 1.98 |
| 6 | 196,132 | 187,620 | 1.91 |
| 7 | 188,308 | 190,728 | 2.03 |
| Week | 1,040,051 | 1,346,397 | 2.59 |

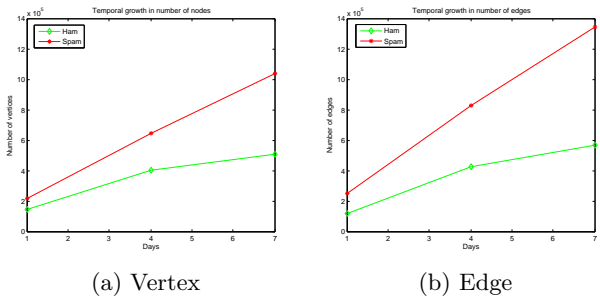

(a) Vertex      (b) Edge

Figure 1: Growth of number of vertices and edges in spam and ham networks after 1 day, 4 days, and the complete week

Tables 3 and 4 illustrate the statistics for ham and spam networks, respectively. As can be seen, the spam networks generated (both for each separate day and after the whole week) are larger than the respective ham networks in terms of number of nodes and edges. Figure 1 shows the evolution of the spam and ham e-mail networks as the number of nodes and edges grow from the first day until the 7th day. As can be seen the growth is almost linear for spam network but not for the ham network. The mean degree for spam networks in each day is also greater than the respective ham network. Note that the spam network only contains the accepted spam e-mails that managed to pass the pre-filtering phase of the receiving mail server.

## 4. ANALYSIS OF SOCIAL PROPERTIES OF THE E-MAIL NETWORKS

In this section we discuss our analysis of the various structural and temporal properties of e-mail networks.

First, the topology of e-mail networks is studied as undirected graphs focusing only on the existence of an e-mail communication between two e-mail addresses. Then the networks are analyzed as directed graphs taking into account the direction of each e-mail transmission. We have investigated the social properties of e-mail networks for the graphs generated both on a day-by-day basis during the course of a week (daily networks), and for the accumulated networks generated after the whole week (weekly networks).

We have looked into the robust measures of network topology that are widely used in order to understand the topology and the structure of different types of networks [8]. Social networks are small world networks with a relatively short path between any two vertices and they are highly clustered. Social networks are also scale-free networks since their degree distribution follows a power-law distribution [1]. Existence of a giant strongly connected component is also another structural property of social networks. In this section we have analyzed these characteristics and their temporal evolution for complete e-mail networks as well as ham and spam networks.

### 4.1 Average Path Length

The "small world phenomenon" often referred to as "six degree of separation" was first studied in pioneering work of Stanley Milgram in 1967. A network exhibits the "small world phenomenon" if any two vertices in the network are likely to be connected through a short sequence of intermediate vertices [35]. In social networks, shortest path length $l$ counts the number of acquaintances in the shortest chain connecting two people [9]. Online social networks [14], www [10, 11, 12], e-mail networks [1], etc. are shown to be small world networks.

In this section, we have calculated the average of $l$ over all pairs of vertices in our ham and spam e-mail networks. We have treated all edges as undirected and have calculate shortest paths only in the giant strongly connected component of the networks which contain a large fraction of vertices of the network where a path exists between any pair of them [9].

Table 5, shows the average shortest path length for daily and weekly ham and spam networks. As can be seen, the ham network usually has smaller average shortest path except for day 7. In day 7 which was a Sunday the number of nodes in the giant strongly connected component of ham network was significantly less than other days. The value of $\langle l \rangle$ depends on the number of nodes in a network [8]. Although the number of nodes in the connected components of ham and spam networks are different (see section 4.4), but it is clear that they are both small world networks.

Table 5: Comparison of average shortest path lenghts in daily ham and spam networks

| Day  | $\langle l_{ham} \rangle$ | $\langle l_{spam} \rangle$ |
|------|------|------|
| 1    | 10.10 | 11.82 |
| 2    | 8.05  | 11.09 |
| 3    | 9.34  | 10.46 |
| 4    | 9.05  | 9.37  |
| 5    | 8.26  | 10.61 |
| 6    | 5.33  | 11.76 |
| 7    | 13.69 | 12.42 |
| Week | 7.70  | 9.09  |

## 4.2 Clustering Coefficient

Watts and Strogatz [9] showed that in addition to a short average path length, small world networks are highly clustered. The existence of short average path lengths between nodes in a small world network is not specific to social networks, even random networks exhibit the same property. But in social networks, the network is fragmented into clusters of individuals with similar characteristics, leading to a relatively high clustering coefficient. The clustering coefficient $C_v$ of a vertex $v$ is given by

$$C_v = \frac{2E_v}{k_v(k_v - 1)}$$

where, $k_v$ is the number of neighbors of $v$, $k_v(k_v - 1)/2$ is the maximum number of edges that can exist between the neighbors of $v$, and $E_v$ is the number of the edges that actually exist (the number of triangles in the network structure).

The clustering coefficient of the network can be calculated by averaging $C_v$ for all nodes in the network.

$$C = \frac{1}{|V|} \sum_{v \in G} C_v$$

Table 6: Comparison of clustering coefficient in daily and weekly e-mail and the respective random networks

| Day  | $C \times 10^{-3}$ | $C_{rand} \times 10^{-6}$ |
|------|------|------|
| 1    | 1.56 | 1.60 |
| 2    | 1.45 | 0.76 |
| 3    | 1.56 | 1.90 |
| 4    | 1.48 | 1.16 |
| 5    | 0.99 | 2.88 |
| 6    | 0.74 | 1.44 |
| 7    | 0.75 | 3.80 |
| Week | 1.68 | 0.33 |

Table 7: Comparison of clustering coefficient in daily and weekly ham and the respective spam networks

| Day  | $C_{ham} \times 10^{-3}$ | $C_{spam} \times 10^{-3}$ |
|------|------|------|
| 1    | 8.40 | 0.92 |
| 2    | 7.49 | 0.66 |
| 3    | 9.99 | 0.83 |
| 4    | 9.33 | 0.80 |
| 5    | 7.31 | 0.37 |
| 6    | 5.66 | 0.39 |
| 7    | 9.53 | 0.44 |
| Week | 9.95 | 1.19 |

The average clustering coefficients for the undirected daily and weekly e-mail networks are shown in Table 6. It can be seen that the average clustering coefficient of each e-mail network is significantly greater than that of a random graph with the same number of vertices and average number of edges per vertex. Note that although we do not see all the communications among internal nodes with each other or among external nodes with each other, the clustering coefficient of our networks is still significantly greater than that of the random graphs. So, as Ebel et al. [1] also observed in their e-mail network which also did not contain the communication of external users with each other, "a high clustering coefficient is a characteristic of e-mail networks".

Since the clustering coefficient in a social network shows to what extent friends of a person are also friends with each other, it is expected that the ham network should have a greater clustering coefficient compared to the spam network [5, 2]. The observed average clustering coefficient for the ham network after the whole week is $C_{ham} = 9.95 \times 10^{-3}$ and for the spam network that only includes delivered spam e-mails is $C_{spam} = 1.19 \times 10^{-3}$. This value is even smaller for rejected and delivered spam network $C_{rejected+spam} = 0.93 \times 10^{-3}$, which confirms the unsocial behavior of spam traffic. Table 7 shows that this difference between ham and spam also holds for the daily networks.

## 4.3 Degree Distribution

The most fundamental structural property of networks is their degree distribution, which is widely used for modeling real networks. Degree distribution of a network is the probability distribution of the node degrees over the whole network. It is a common property of many real networks such as the Internet [17], the World Wide Web[10, 12], phone call graphs [13], and online social networks [14] to exhibit a power-law degree distribution [15, 7].

In a power-law distribution, the degree of a node is proportional to the fraction of nodes in the network with degree $k$, $n(k)$, to the power of a constant, $\gamma$,

$$n(k) \propto k^{-\gamma}$$

Networks exhibiting such degree distribution are called scale-free [15, 17].

Ebel et al. showed in [1] that e-mail networks are scale-free, since the degree of the nodes follows a power-law distribution ($n(k) \propto k^{-1.81}$). The e-mail network used by Ebel et al. was limited to log files of the mail server of their university, and thus biased to local behavior. Our data, on the other hand, is not limited to a local domain, and is orders of magnitude larger than their data set, although it was collected during a shorter period of time.

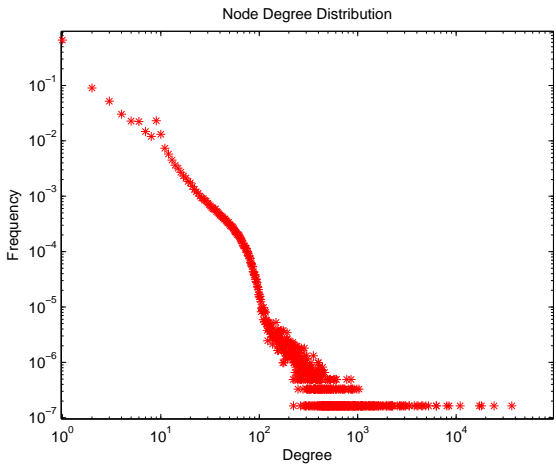

Figure 2: Degree distribution of the complete weekly e-mail network

Figure 2 shows the log-log plot of the degree distribution of the complete undirected weekly e-mail network. The $x$-axis denotes the degree of a node, and the $y$-axis denotes the fraction of nodes in the network with that degree. As a surprise, it can be seen that the e-mail network is not scale-free, since the degree distribution does not follow a power-law distribution. This observation is in contrast with previous works that observed scale-free behavior on their datasets [1, 2, 5, 6]. Note that the e-mail network studied here is significantly larger and more complete than the previous studies.

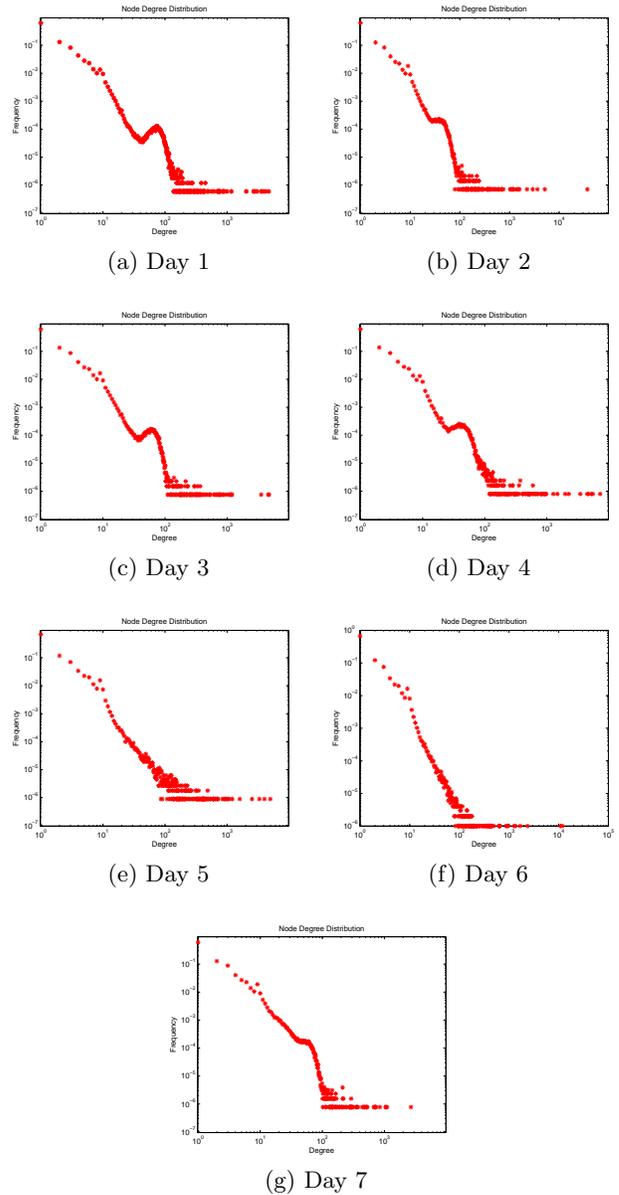

(a) Day 1  (b) Day 2
(c) Day 3  (d) Day 4
(e) Day 5  (f) Day 6
(g) Day 7

Figure 3: Degree distribution of undirected daily e-mail networks

When looking at the degree distribution of the daily e-mail networks (Figure 3) in the first four days and in the seventh day, the deviation from power-law distribution is obvious. For the fifth and sixth days (which were a Friday and a Saturday, respectively) the degree distribution is closer to a power-law distribution but still does not follow the power-law in all points.

We suspect that this deviation from power-law distribution is caused by the large number of unsolicited communications on the Internet. In order to verify this, we have generated a graph including only the legitimate e-mail (ham) transmissions (according to scores generated by SpamAssassin). As shown in Figure 4, the ham network is actually a scale-free network and its degree distribution obeys a power law $n(k) \propto k^{-2.4}$ (Figure 5). This means that the e-mail network of legitimate users represents the same characteristics as other social networks.

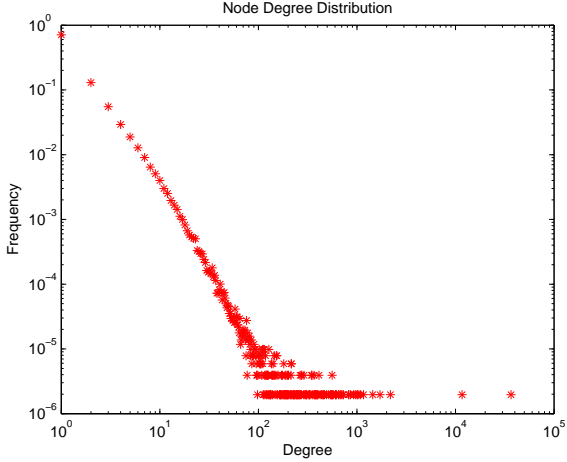

Figure 4: Degree distribution of undirected weekly ham network

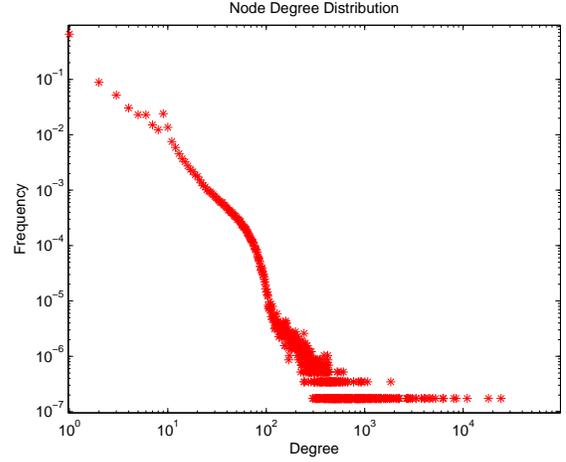

Figure 6: Degree distribution of weekly spam network (including both delivered and rejected spam e-mails)

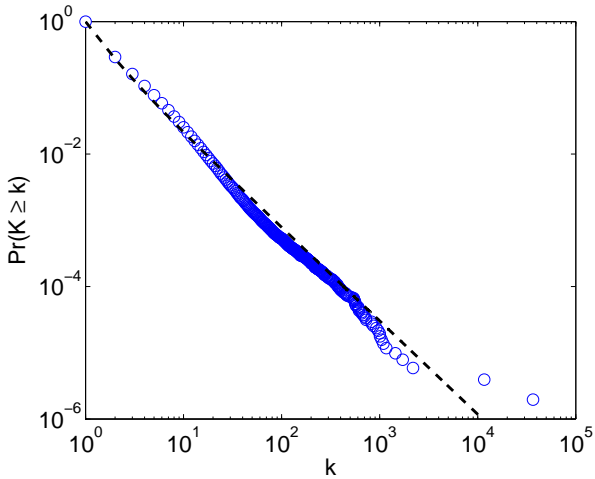

Figure 5: Degree distribution of undirected weekly ham network that exhibits a power-law distribution with estimated exponent $\gamma = 2.4$ ($n(k) \propto k^{-2.4}$)

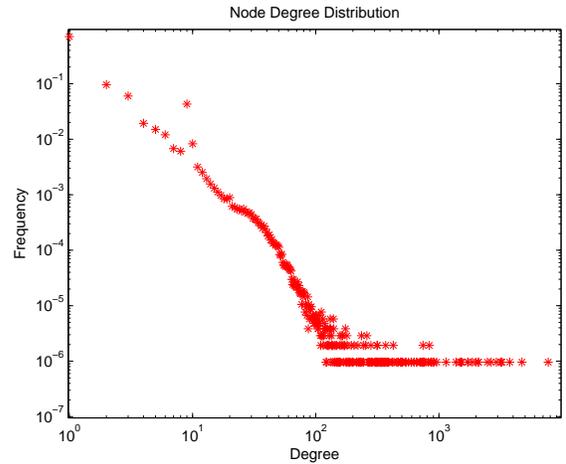

Figure 7: Degree distribution of undirected weekly spam network (including only delivered spam e-mails)

It is known that the vast majority of spam e-mails are automatically generated, so we expect that they don't show the same social behavior as human-generated e-mail traffic. To investigate the difference, we have created a network based on e-mail traffic that were either spam or were rejected by the receiving mail servers (by pre-filtering strategies such as blacklisting). Figure 6 shows the vertex connectivity of this network. It can be clearly observed that the delivered and rejected spam e-mail network does not exhibit a power-law degree distribution.

Unfortunately, it is not possible to accurately separate rejected e-mail traffic due to greylisting, temporary errors, and miss-configurations, from those that are rejected due to IP blacklisting or other anti-spam strategies (see section 3). In order to examine the social characteristics of spam more accurately, we have constructed another e-mail network that only contains delivered spam e-mails that successfully passed the pre-filters (according to the SpamAssassin scores), and studied the degree distribution of this network. Figure 7 clearly indicates that the spam network does not show scale-free behavior over a large range.

To further investigate the existence of a scale-free de-gree distribution, the e-mail network was examined as a directed graph. Directed networks are characterized by two degree distributions: the distribution of outgoing edges which is the probability that a user have sent $k$ e-mails, and the distribution of incoming edges which is the probability that $k$ users have sent e-mail to a certain e-mail address. Figure 8 shows the daily as well as the weekly in-degree and out-degree distributions. In-degree of a node is the number of nodes it have received e-mail from, and out-degree of a node denotes the number of e-mails it has sent to others. Again, the figures show that the degree distribution varies per day. It can be seen that the distribution of out-degree deviates much more from the power-law distribution compared to the in-degree. In [1], the authors observed that both the in-degree and out-degree of e-mail networks follow the power-law distribution with different exponents. However, the out-degree distribution for only the internal users in their dataset was not showing a scale-free behavior. This shows that the study of internal communications on its own is probably not enough for observing the real structure of the networks.

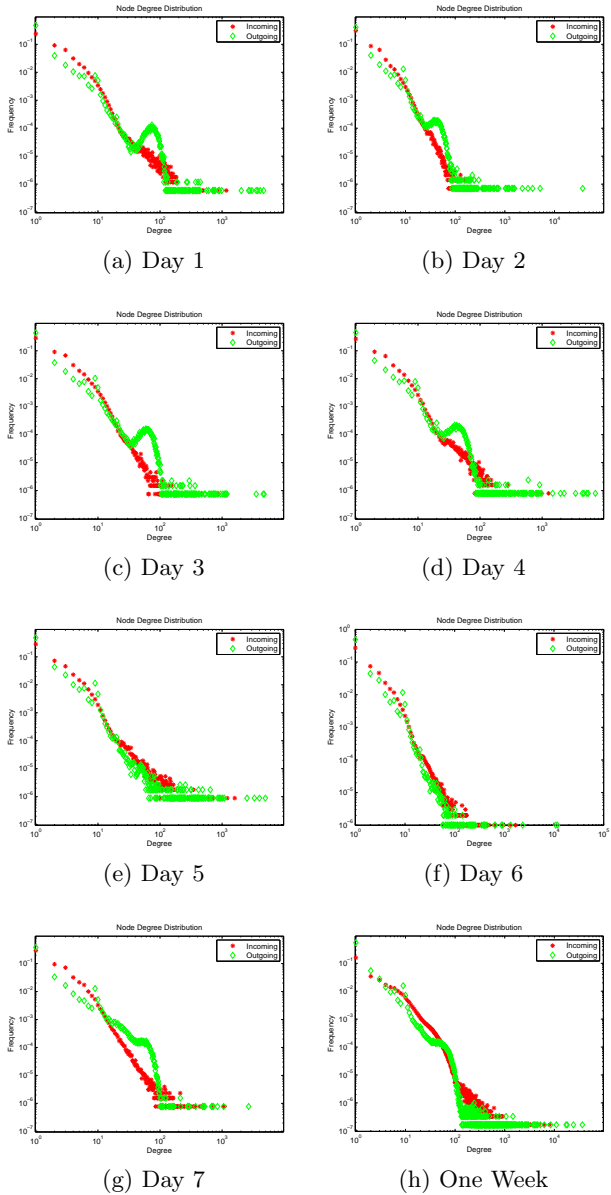

Figure 8: In-degree (incoming) and out-degree (outgoing) distribution of directed daily e-mail networks and the complete weekly e-mail network

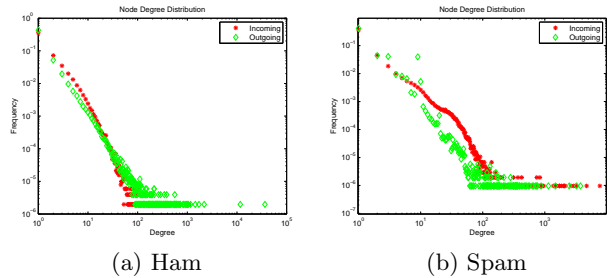

Figure 9: In-degree (incoming) and out-degree (outgoing) distribution of weekly ham and spam networks

Plotting the in- and out-degree distribution for weekly ham and spam in Figure 9 confirms that both the in- and out-degree distributions for legitimate e-mail transmissions follow a power-law distribution (with different exponents). Hence, ham network has scale free characteristics of other social networks. The deviation from power-law, especially observed in the out-degree distribution of spam network is caused by the non-social behavior of spammers in automatically sending spam.

It is also interesting to see how these characteristics evolve when aggregating data over a longer period of time. Therefore, we have studied the changes in vertex connectivity as the network grows over time. Again, we have first studied e-mail networks as undirected graphs and then as directed graphs. Figure 10 shows the temporal evolution of degree distribution of the undirected graph from the first day until the 7th day. It seems that by accumulating more e-mail communications in the network, the degree distribution gets closer to a scale-free network. This also holds for the in-degree distribution of the directed complete e-mail network (Figure 11), but it is not the case for the out-degree distribution.

The variation of the in- and out-degree distribution for the ham network is shown in Figure 12. The in- and out-degree distributions are quite similar for the networks generated after 4 days and 7 days of measurements. This also means that our short duration of measurements (compared to Ebel et al. [1]) is not affecting the existence of power-law degree distribution and the value of its exponent. This is not the case for the spam network (Figure 13) which confirms that legitimate traffic is similar to other social networks, but spam is different since it is not social.

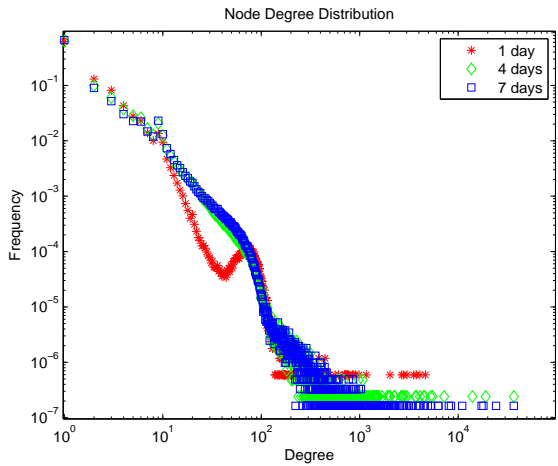

Figure 10: Temporal variation of node degree distribution of the undirected complete e-mail network

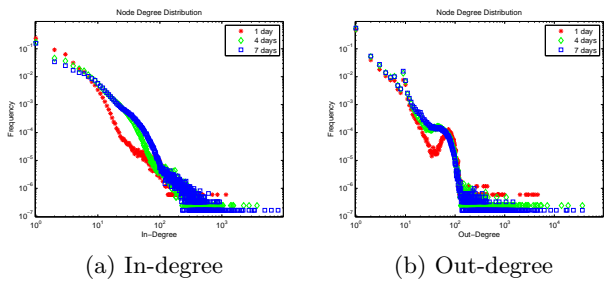

Figure 11: Temporal variation of in-degree (incoming) and out-degree (outgoing) distribution of directed complete e-mail network

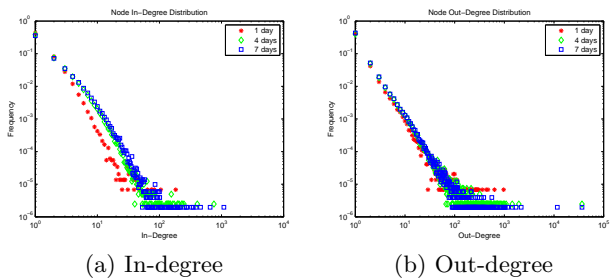

Figure 12: Temporal variation is in-degree (incoming) and out-degree (outgoing) distribution of directed ham network

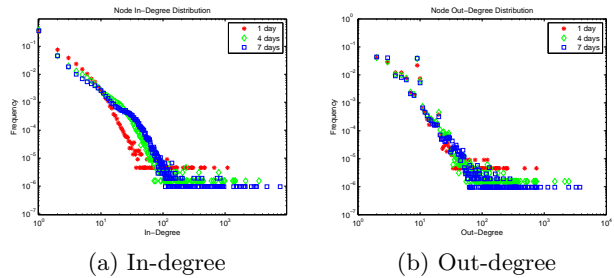

Figure 13: Temporal variation is in-degree (incoming) and out-degree (outgoing) distribution of directed spam network

### 4.4 Strongly Connected Components

Another important structural property of complex networks, including social networks, which is also used for generating realistic network models is the presence of a giant strongly connected component [13, 1]. A strongly connected component (SCC) is a subset of vertices of a network that can both reach and be reached from any other vertex in the same set. A giant SCC (GSCC) contains a significant fraction of the vertices in the network.

This property was mainly studied in the literature to investigate the spread of infections such as the propagation of e-mail-attached viruses that can forward themselves to e-mail addresses of users' address books [1, 23]. This property was also used in analyzing modeling of real networks such as the web graph [12], online social networks [14] and the phone call graphs [13].

Table 8: Percentage of total number of nodes in the GSCC of undirected daily and weekly e-mail, ham and spam networks

| Day | E-mail | Ham | Spam |
| --- | --- | --- | --- |
| 1 | 64.98% | 29.84% | 50.06% |
| 2 | 65.01% | 49.67% | 27.75% |
| 3 | 66.17% | 33.00% | 35.00% |
| 4 | 62.77% | 34.92% | 43.13% |
| 5 | 41.81% | 32.97% | 29.99% |
| 6 | 41.84% | 31.69% | 24.56% |
| 7 | 66.17% | 10.78% | 30.61% |
| Week | 81.08% | 68.60% | 51.14% |

The undirected weekly e-mail network studied in this paper is comprised from 326,473 separate strongly connected components with a GSCC containing 4,943,470 nodes which is 81.08% of the total number of nodes in the network. The second largest SCC of this e-mail network contains only 766 nodes which is approximately 6,400 times smaller than the GSCC. For each individual day, the e-mail networks also have a GSCC with different fraction of nodes ranging from 41% to 66% of the total number of nodes in the network. All the weekly and daily ham and spam networks also have a number of disjoint components with a giant strongly connected component. The GSCC of the weekly ham network contains 68% and the GSCC of the spam network contains 51% of the total number of nodes in the respective network. The daily ham and spam networks have smaller percentage of nodes in their GSCCs tahn the weekly networks as can be seen in Table 8. Considering all the rejected and delivered spam traffic together, the GSCC of the undirected weekly network contains $4,642,015$ nodes which is 80.44% of the total number of nodes in its network. Note that the components with SCC of size 1, show the self-loops in the graph that are e-mails sent from one person to herself.

The distribution of the size of the SCCs for the undirected weekly e-mail network is shown in Figure 14. The $x$-axis denotes the size of each SCC of the network, and the $y$-axis denotes the fraction of nodes in that SCC. Both axes are plotted in logarithmic scale. It can be seen that the GSCC of the network is orders of magnitude larger than other strongly connected components of the network.

Figure 15 illustrates that a GSCC similarly exists in the weekly ham, spam, and rejected plus delivered spam networks. However, the shapes of the distributions are not similar. The distribution for the ham network follows a power-law distribution, but the shape of the distribution for the spam network and the rejected plus delivered spam network are slightly different and some outliers are present in the distributions. Therefore, the distribution of the size of SCCs of the complete e-mail network does not follow the power-law distribution (Figure 14).

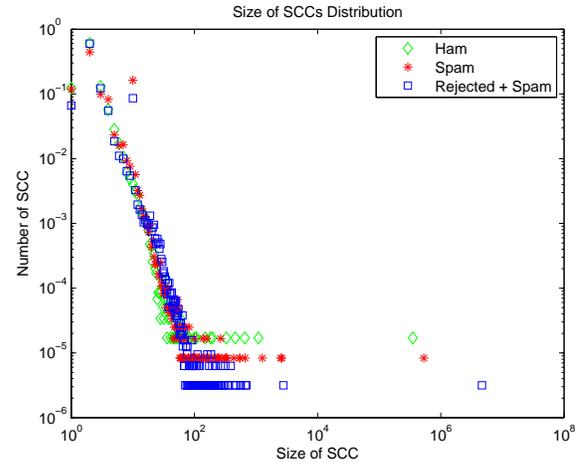

Figure 15: Distribution of the size of strongly connected components for the undirected weekly ham, spam and rejected plus delivered spam e-mail networks

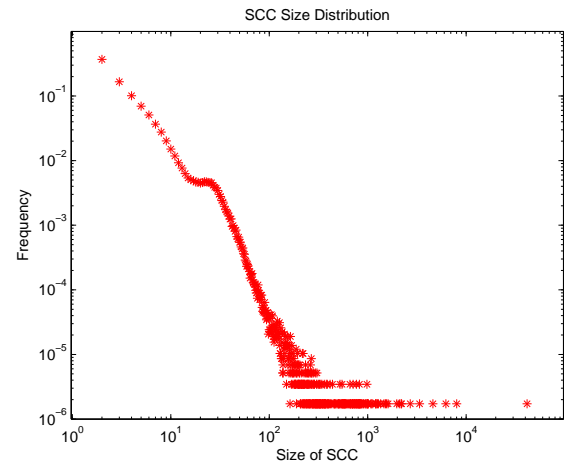

Figure 16: Distribution of the size of strongly connected components for the directed weekly e-mail network

The directed weekly e-mail network studied here contains 580,112 separated connected components and its largest SCC contains only 41,593 vertices, which is around 0.68% of the total number of nodes in the network, i.e. it is not a GSCC. Figure 16 demonstrates the SCC distribution for the directed weekly e-mail network in logarithmic scale. The $x$-axis denotes the size of each SCC of the network, and the $y$-axis shows the fraction of nodes in that SCC.

The largest SCC of the ham network contains 3.28% of the total number of nodes in the graph which is not a large fraction of nodes. However, this value is still much larger than the largest SCC of the spam network which contains only 0.74%, and the largest SCC of the rejected and delivered spam network that includes only 0.14% of the nodes in the network. Figure 17 illustrates

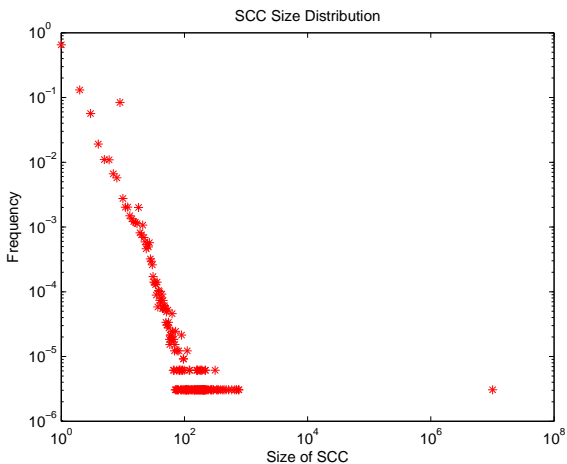

Figure 14: Distribution of the size of strongly connected components for the undirected weekly e-mail network

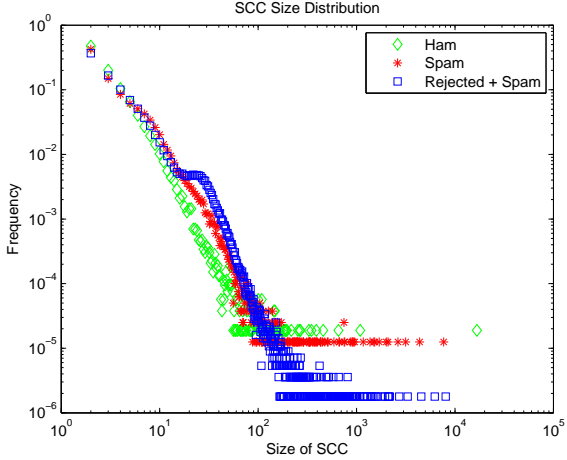

Figure 17: Distribution of the size of SCCs for the directed ham, spam, and rejected plus delivered spam networks generated after the whole week

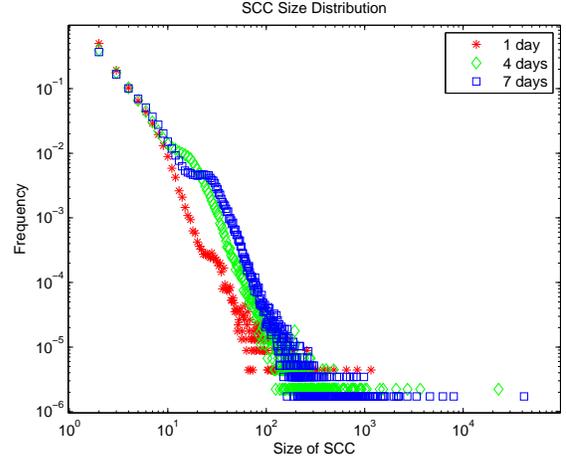

Figure 19: Temporal variation in the distribution of the size of SCCs for the directed e-mail network generated after 1 day, 4 days, and the complete week

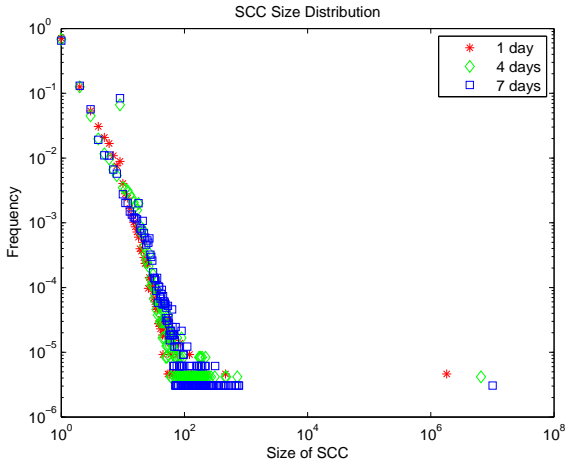

Figure 18: Temporal variation in the distribution of the size of SCCs for the undirected e-mail network generated after 1 day, 4 days, and the complete week

that a GSCC does not exist in the directed spam, and the rejected plus delivered spam networks.

The small size of the largest SCC of the studied networks compared to the previous studies that examined phone call graphs [13] and the web [12] (bothe of them have a GSCC containing around 28% of the total number of nodes in their respective directed networks) can be due to measurement shortcomings. For example, due to routing policies we see less traffic on the outgoing direction compared to the incoming direction of the measured backbone link.

Figure 17 also shows that the rejected spam network have a different distribution shape for the size of its SCCs compared to the ham network. This deviation which similarly exists in the undirected network is in contrast with the SCC size distribution of the web graph [12] and phone call graphs [13] and the ham network where the distributions were obeying the power-law distribution.

The temporal variation in the distribution of size of SSCs in Figure 19 and Figure 18 shows that by increasing the number of nodes in the network, the size and the number of the connected components increases in both directed and undirected e-mail networks. In the undirected network, the GSCC size increases from 64.98% in the first day to 81.01% after 4 days and reaches 81.08% after the whole week. For the ham network, the size of the GSCC also increases from 29.84% to 64.50% and reaches 68.6% after the whole week. However, the size of the GSCC of the spam network seems to be little changing from 50.06% to 52.31% and 51.14%.

According to our findings, by increasing the size of the network, the deviation of the distribution of the size of SCCs of the networks from the respective power-law distribution becomes clearer, especially for the directed networks case.

## 5. DISCUSSION

In this section the observed structural and temporal characteristics of the e-mail networks examined in this paper are discussed.

The e-mail networks studied in this paper were generated from SMTP traffic captured on an Internet backbone link. The traffic we have captured contains a good mix of traffic from several universities, student dormitories, corporate users, and home users and thus is more complete compared to previous attempts to study the topology and structure of e-mail networks [1, 23, 2, 6].

We have observed that e-mail networks are "small world" networks with short average path lengths and high clustering coefficients. Our results indicate that the clustering coefficient is a discriminative characteristic for ham and spam networks. Networks of legitimate e-mails have higher clustering coefficient values confirming the results in [5]. So this property can be deployed in different anti-spam tools. Average shortest path length can also be used to distinguish ham from spam traffic, but its calculation is computationally prohibitive (depends on the number of nodes in the network ($O(log\ n^3)$)) and thus not reasonable for large scale networks.

We have also shown that despite the common belief that e-mail networks are scale-free [1, 2, 3, 4, 6, 5], the e-mail network generated from real network traffic is not scale-free. We have shown that only legitimate e-mail traffic exhibit scale-free behavior, and the deviation of the degree distribution from power-law distribution is mainly caused by spam traffic, especially spam e-mails sent by known spammers that were rejected during the pre-filtering process by the receiving mail servers. We conjecture that Ebel et al. [1] did not observe this in their e-mail network because their data was more local with fewer spammers or the log files only contained delivered e-mails. Gomes et al. [2] also studied the e-mail transmissions limited to one university department, and although their data contained rejected and spam traffic they did not observe non-power-law behavior in the degree distributions. We think it can be due to the method they used to generate their e-mail networks by classifying e-mail addresses as spammer and non-spammer. Boykin et al. [5] also observed that spam e-mail has a power-law distribution but with an exponent greater than that of legitimate e-mails. However, their e-mail network was limited to e-mail communications of a single user.

We have also observed that the in- and out-degree distributions are different when looking into directed daily and weekly e-mail networks. The out-degree clearly does not follow a power-law distribution. This is due to the non-social behavior of spammers in automatically sending large amount of spam.

With time as the networks grow larger and more connected, it is expected that the degree distribution will get closer to a power-law distribution, even for spam networks. Therefore, it turns out that it is not necessary to collect data for long periods of time to be able to detect the existence of spammers in a network. In other words, if we want to take advantage of this distinguishing difference between ham and spam, we should construct e-mail networks with a shorter time window. In addition, by creating the network from a shorter period of time the number of nodes and edges in the network would be smaller and the analysis of the data becomes easy and fast.

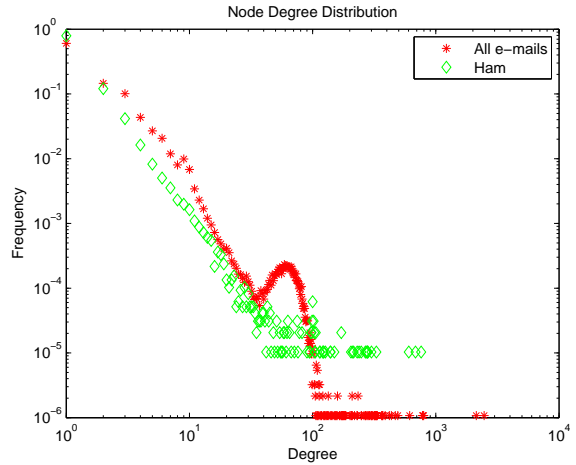

Figure 20: Comparison of degree distribution for complete undirected e-mail network and ham network generated from 12-hour measurement

The selection of the size of the time window can depend on the network being measured. During the daytime and working hours more ham is exchanged, although spam activity stays almost constant during the 24 hours [19]. All these factors could be taken into account when selecting a time interval for capturing the traffic and generating the e-mail networks.

In order to verify that the study of degree distribution is still useful with shorter time windows, we have generated an e-mail network from e-mail traffic captured in a 12-hour window on Wednesday (day 3) from 6 am to 6 pm (working hours). The generated network has $926,845$ vertices and $1,229,718$ edges, and as Figure 20 shows, the degree distribution for the whole e-mail network clearly deviates from a power-law distribution. However as the figure shows, the ham network generated from this dataset still shows a scale free behavior. Therefore, it seems that for our dataset even a 12-hour time window (as an alternative to 24 hours or a complete week) during working days could clearly identify the existence of spamming activities.

Finally, we have shown that a giant strongly connected component (GSCC), which usually exists in social networks, is similarly present in both the undirected spam and ham networks. However, the shape of the distribution of the size of SCCs differs for ham and spam, especially when rejected traffic is also considered as spam. This distribution for the ham network obeys a power-law distribution, but this is not the case for the rejected and the spam traffic. Also as the networks evolve over time, the ham network becomes more connected but the percentage of nodes in the GSCC of the spam network changes a little.

## 6. CONCLUSIONS

In order to better understand the social behavior and dynamics of e-mail users, we have performed a study of the structural properties of e-mail networks. Understanding the topology of e-mail networks and fundamental social properties of e-mail traffic can lead to a better insight into the e-mail communication of legitimate users and new ways to detect spamming activities on the Internet.

Our analysis on the e-mail network generated from SMTP traffic on an Internet backbone link revealed that contrary to previous studies, which were based on very limited datasets, connectivity of nodes in e-mail networks does not show a power-law behavior. This deviation from scale-free characteristic found in other social networks is caused by the extensive number of spam present on the Internet.

Our study of the social network dynamics and characteristics of spam and legitimate e-mail traffic indicates that there are similarities and differences. The most significant difference is that legitimate e-mail traffic is social while non-legitimate traffic such as spam does not show social behavior. The differences revealed in this study could potentially be used to compliment current anti-spam tools and lead to new methods to detect spamming nodes on network-level closer to the source.

## 7. ACKNOWLEDGMENTS

This work was supported by .SE – The Internet Infrastructure Foundation.